# Crowds in front of bottlenecks at entrances from the perspective of physics and social psychology

Juliane Adrian[1], Armin Seyfried[1,2], Anna Sieben[1,3]

   (1) Institute for Advanced Simulation, Forschungszentrum Jülich, Germany
   (2) Faculty of Architecture and Civil Engineering, University of Wuppertal, Germany
   (3) Faculty of Social Sciences, Ruhr University Bochum, Germany

## Abstract

This article presents an interdisciplinary study of physical and social psychological effects on crowd dynamics based on a series of bottleneck experiments. Bottlenecks are of particular interest for applications such as crowd management and design of emergency routes because they limit the performance of a facility. In addition to previous work on the dynamics within the bottleneck, this study focuses on the dynamics in front of the bottleneck, more specifically, at entrances.

The experimental setup simulates an entrance scenario to a concert consisting of an entrance gate (serving as bottleneck) and a corridor formed by barriers. The parameters examined are the corridor width, degree of motivation and priming of the social norm of queuing. The analysis is based on head trajectories and questionnaires.

We show that the density of persons per square metre depends on motivation and also increases continuously with increasing corridor width meaning that a density reduction can be achieved by a reduction of space. In comparison to other corridor widths observed, the narrowest corridor is rated as being fairer, more comfortable and as showing less unfair behaviour. Pushing behaviour is seen as ambivalent: it is rated as unfair *and* listed as a strategy for faster access.

### Keywords

# Introduction

Bottlenecks occur for a variety of reasons and fulfil different functions. When an event or protected area is entered, bottlenecks are used for access or ticket control and security checks. In buildings, trains and buses, doors or narrow corridors or aisles are necessary for static, technical or economic reasons. As they limit the performance of a facility or a process and cause congestions, they are of special interest for applications such as crowd management, design of emergency routes, and science. Flow and capacity of bottlenecks depend on various factors. These can be categorised according to factors related to the spatial structure of the bottleneck and factors related to physiological and psychological characteristics of the people passing through the bottleneck as well as social dynamics in the crowd. Observations on the effects of spatial structures concern, for example, the width and length of the bottleneck [1–9]. A review of these effects is given, for instance, by [10] and [11]. Physiological and psychological factors may involve stress, social norms, identity or motivation [4,12–18]. Obviously, the dynamics within the bottleneck are related to the dynamics in front of it. While most studies focus on the flow through the bottleneck, only a few examine the situation in front of it [18–21]. This interdisciplinary paper addresses physical as well as social psychological effects on crowd dynamics, both within the bottleneck and in front of it. It is based on a previous experimental and questionnaire study [18] but more systematically examines corridor width, motivation and social norms.

## Motivation and flow

The effect of motivation in entrance situations is assumed to be straightforward: the more participants are motivated to access fast, the denser the situation becomes and the faster the initial velocity is. The motivation to 'access fast' can be further subdivided into the motivation to be one of the first (to be faster than the others) or to require little total time (regardless of whether one is faster or slower than the others) or to be fast together (total time for everyone accessing) and thus results in more competitive or cooperative behaviour. In general, motivation is triggered by a reward.

In this context, a prominent phenomenon that has been described and modelled in many works is the 'faster-is-slower' effect [13, 22 and 16]. This term is used to describe the fact that the flow through a bottleneck is reduced when the motivation of the individuals involved is high. The reduction of the flow is caused by temporal clogs limiting the flow. The empirical evidence for this phenomenon is, however, sparse. It is important to note that properties of the spatial structure of the bottleneck show complex correlations with psychological factors clearly confining the occurrence of the 'faster-is-slower' effect: experimental studies in which the motivation and the width of the bottleneck were varied have shown that the flow increases with the motivation and the 'faster-is-slower' effect occurs only at very narrow bottlenecks ($w < 0,7$ m) and very high motivation [4,12 and 15–17].

The degree of motivation changes the dynamics of a crowd within the bottleneck as well as in front of it. In [23], the authors examine the polarisation of velocity within a crowd in a bottleneck experiment. They observe that high motivation can lead to temporary collective transversal movement which in turn increases the risk of tripping and falling.

## Queuing and spatial structure

In order to manage entrance situations and to prevent very dense or uncomfortable situations, the social system of queues has been established in many cultural contexts [e.g. 24 and 25]. If an entrance procedure is clearly organised as a queue, relatively well-defined social norms apply (for an overview, see [26]): queue jumping or pushing in is not allowed [24 and 25]; except for those who ask whether they are allowed to cut in line, people are supposed to wait one after the other and they are expected to keep a comfortable personal distance.

Social norms and perceived fairness of an entrance situation are interconnected and it is possible to evaluate different aspects of entrance procedures as just/unjust: what is most important here is the FIFO principle (first in, first out) [27]. Again, fairness is best defined and easily observable if an entrance is organised as a queue. In [28], the authors have shown that participants evaluate multiple parallel queues as less just and less pleasant than single queues.

Some entrance procedures are clearly designated as queues, for instance, with signs. But more often, queues are established by convention: if people expect a queue in a certain context, they are likely to form a queue. One such example is a check-in area at an airport. Queuing conventions can be very specific for a local or cultural context. Sometimes, the formation of a queue follows the initiative of the first arrivals who have established the queue. In other situations, queues are enforced by spatial structures: corridors for queues allow only very few people to stand side by side and often they have to stand in single file. Queues at airports in front of passport control are a prime example of this.

The transition from an ordered entrance procedure to an unordered one is, to the best of our knowledge, only studied in [18]. This study focuses on the spatial structure of the barriers in front of a bottleneck. It was found that the density is significantly lower if participants are guided by a corridor in comparison to an open space setting in which participants form a semi-circle. On the basis of a post hoc questionnaire study, queuing norms were identified as one potential cause of this effect. It is not known how wide a corridor can be in order to be recognised as a structure for a queue and this is addressed in the present study.

If the entrance scenario is not recognised as a queue, it becomes normatively more unclear. Previous research [18] has shown that participants associate an entrance scenario in which a crowd forms a semi-circle in front of the entrance with principles such as 'right of the stronger' or 'the stronger wins'. Some participants argue that in this entrance scenario no rules apply. At the same time, participants were ambivalent about whether pushing and shoving are forbidden or allowed.

Furthermore, humans act differently in entrance situations, depending on what strategy they use for accessing. In our previous study, the strategies most often mentioned were pushing and shoving, using elbows, using gaps and choosing the left or right side for walking. These strategies are more likely to be used if participants think that they are able to contribute to accessing faster. Therefore, we hypothesise that entrance procedures in which participants list strategies such as pushing or using gaps and think that these are able to contribute show higher densities. Spatial structure such as corridor widths presumably make some strategies appear more promising than others.

## Design and hypotheses of the actual experiments

The present study looks at the relationships between corridor width, motivation, social norms, density, waiting times and flow. It systematically compares width for straight corridors and, in addition, varies high and low motivation and the social norm of queuing via a priming procedure. We hypothesise that
   a) density increases with corridor width
   b) runs with high motivation result in higher densities than runs with low motivation
   c) priming the social norm of queuing decreases density.

Thereby, the present study responds to the following two limitations of the earlier study [18]: First, the corridor led the participants perpendicularly towards the entrance and the open space suggested a straight movement from the individual starting point. To avoid such a simultaneous variation of width and direction in the present study only straight corridors are built. Second, only two different spatial structures were explored, a two-metre wide corridor and an open space. Now, five different widths are used.

The questionnaire study looks at the relationships between corridor width and social norms, perceived fairness, comfort, perceived inappropriate behaviour and strategies for accessing. Unlike the precursor study (in which participants watched videos of the entrance procedure and filled in the questionnaire afterwards), this time the questionnaire is answered by participants who walked through the entrance structure themselves. We hypothesise that
   d) narrower corridors are perceived as fairer, more comfortable and that participants observe less inappropriate behaviour here, and
   e) that in narrower corridors fewer participants think they are able to contribute to accessing faster and fewer participants say that they themselves were pushing.

The present manuscript focuses on the analysis of the experimental data. A comparison between the experimental findings with simulated data is done by [29].

# Methods

## Experimental setup

The experimental setup (see Figure 1) consisted of barrier systems, as is common for concerts or other events. A typical entrance gate with a passage width of 0.5 m served as a bottleneck. The corridor leading straight to the entrance gate had a certain width $w$ which was varied between 1.2 m and 5.6 m.

## Participants

For each set of experiments (including two experimental runs, one with high, one with low motivation), participants were recruited in university lectures in order to avoid multiple participation as far as possible. The venue for the experiments was located between two lecture halls. The research team introduced themselves and the experimental procedure to the students during the lectures and asked them to participate. We collected written informed consent from all students who were willing to take part. They were rewarded with a voucher for the university dining hall (worth approximately 5 euros). In total, $N = 479$ students participated, with 46.9% identifying as female and 52.2% as male. The ad hoc recruiting strategy resulted in varying numbers of participants for each experimental set.

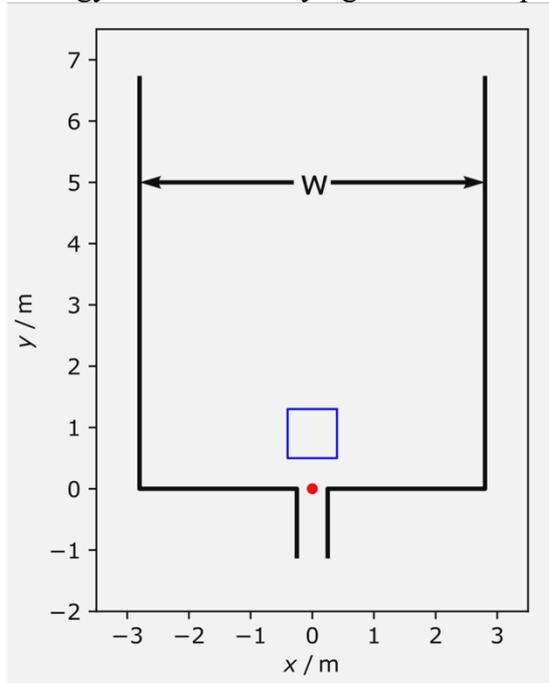

*Figure 1: Sketch of the experimental setup; w is the corridor width, the red dot is the target at the entrance of the gate, and the blue square indicates the measurement area.*

## Procedure

From the lecture hall, participants were guided to a cloakroom where they had to store their jackets and bags and were given orange hats. As a group they were then placed loosely into the experimental structure. The following instructions for the first run (high motivation) were read aloud to them (translated from German): "Imagine you are on your way to a concert by your favourite artist. You know that at the back you can hardly see anything at all or only the video screen. You absolutely want to be standing next to the stage and therefore want to access the concert as fast as possible. After a signal, we will open the entrance." Then, after a "Go!" signal, the entrance was opened and participants left the experimental structure. Afterwards, they were loosely placed next to the experimental setup and then again placed into the structure as a group. The following second instructions (low motivation) were read aloud to them (translated from German): "Thank you very much! Please imagine again that you are on your way to a concert by your favourite artist. This time you know that everyone will have good view. Still, you would like to access the concert quickly." Again, the entrance was opened and the participants walked out of the structure. They were then guided back to the cloakroom where they were first given the questionnaire and asked to return the hat. The reward was given to students together with their jacket and bags.

## Priming procedure

We used a priming activation scheme [30] in order to make social norms of queuing more salient in one experimental condition and crowding behaviour in the other. Queuing or crowding were primed at two points. First, during the introduction of the research field during the lecture. Students were either told that "we are researching crowded situations such as spontaneous queues at airports" or "we are researching crowded situations such as large crowds in front of entrances". Second, a picture was included on the informed consent form which either showed orderly queues in an airport setting or a dense crowd forming a semi-circle in front of an escalator.

## Overview of experimental runs

An overview of all experimental parameters including corridor width $w$, priming $P$ (c = crowding, q = queuing), degree of motivation $h$ (hi=high, lo=low), number of participants $N$ and percentage of female participants is given in Table 1.

*Table 1: Overview of the experimental runs.*

| Run | Width $w$ / m | Priming $P$ | Motivation $h$ | N | % female |
|---|---|---|---|---|---|
| 030 / 040 | 5.6 | c | hi / lo | 75 | 84.0 |
| 050 / 060 | 4.5 | c | hi / lo | 42 | 45.2 |
| 070 / 080 | 2.3 | c | hi / lo | 20 | 30.0 |
| 090 / 100 | 1.2 | c | hi / lo | 24 | 37.5 |
| 110 / 120 | 1.2 | c | hi / lo | 63 | 87.3 |
| 150 / 160 | 5.6 | c | hi / lo | 57 | 29.8 |
| 170 / 180 | 1.2 | q | hi / lo | 25 | 64.0 |
| 190 / 200 | 3.4 | q | hi / lo | 22 | 50.0 |
| 230 / 240 | 2.3 | q | hi / lo | 42 | 16.7 |
| 200 / 260 | 4.5 | q | hi / lo | 42 | 31.0 |
| 270 / 280 | 3.4 | c | hi / lo | 67 | 13.4 |

## Trajectories

Figure 2 shows a screenshot of the video sequences at time $t$ = 5 s after the starting signal for a run with the smallest corridor width (Run 110) and for a run with the largest corridor width (Run 030).

The obligatory orange hats are used for the automatic detection and tracking of the individual head positions based on the video sequences resulting in head trajectories. The automatic extraction of the trajectories was performed according to [31] including the correction of video recordings in order to compensate lens distortion. These trajectories serve as a basis for measuring density, velocity, waiting time and target distance.

As examples, the automatically extracted individual head trajectories of Runs 110 and 030 are shown in Figure 3. In Run 030, a run with the widest corridor, an initial contraction and a semi-circle arrangement in front of the bottleneck can be presumed.

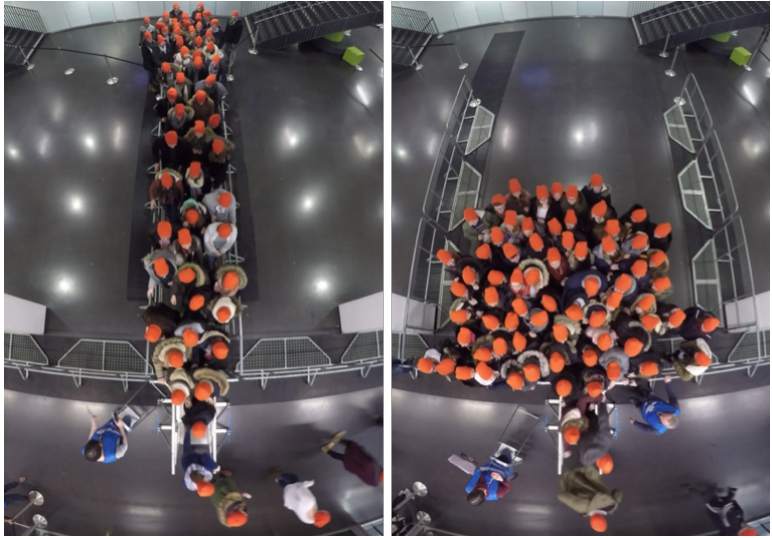

*Figure 2: Screenshots of the video sequences at time t = 5 s for Run 110 with w = 1.2 m (left) and Run 030 with w = 5.6 m (right).*

It should be mentioned that the raw trajectories were corrected before further data processing was performed because several people were leaning their heads over the barriers due to a barrier height of only 1.15 m. This leads to head positions outside the given geometry causing problems in the determination of densities. Therefore, all head positions were corrected to lie inside the geometry. This was realized by displacing points of trajectories that lie too close to or outside of the geometry at least 0.1 m away from the boundary. We are convinced that this procedure is reasonable since the lower body is located inside the geometry as is the corrected head position.

In the following, we will focus only on the analysis of trajectories from experimental sets with more than 40 participants because the duration of runs with fewer participants was too short to derive meaningful mean physical quantities.

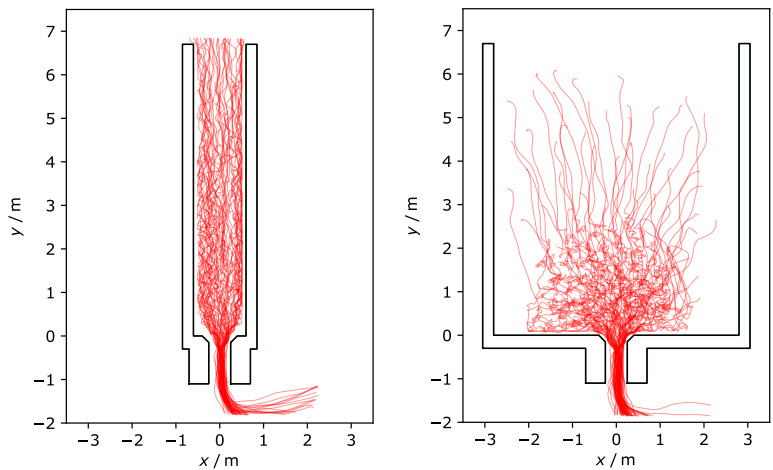

*Figure 3: Trajectories of Run 110 with w = 1.2 m (left) and Run 030 with w = 5.6 m (right).*

## Questionnaire

The questionnaire was administered after the two sequenced runs. Participants were asked to evaluate the first run (high motivation) only. The questionnaire contains six items (translated from German): 1) Did rules apply to the entrance procedure? (yes/no), If so, what rules? (open-ended question), 2) How fair was the entrance procedure? (six-point scale, 1 = very unjust, 6 = very just), 3) How comfortable did you feel? (six-point scale, 1 = very uncomfortable, 6 = very comfortable), 4) Did you observe anyone behaving inappropriately or unfairly? (six-point scale, 1 = no one, 6 = everyone), What kind of inappropriate behaviour did you observe? (open-ended question) 5) Could you contribute to accessing faster? (yes/no), If so, what kind of strategies did you use? (open-ended question), If not, why not? (open-ended question), 6) Did you yourself push? (yes/no), Why? (open-ended question) Respondents were also asked to give their age and gender. The questionnaire is completely anonymous; data can only be assigned to the experimental run, not to individuals.

For perceived fairness, comfort and observed unfair behaviour, a two-way repeated measures analysis of variance (width x priming) was conducted. Chi-squared tests were used for nominal items. All statistical analyses were performed with SPSS for Mac (version 25). The significance level is set at p < .05.

# Results: trajectories

## Qualitative observation and trajectories

The screenshots in Figure 3 for $t = 5$ s represent runs with the narrowest and widest corridors. These runs will be our prime examples, since they show different characteristics. Around $t = 5$ s, the highest density occurs, as will be seen in the following sections. The screenshot of Run 110 indicates that the narrow corridor forces the participants into a queue with two or three individuals side by side. The line of sight of all participants and, consequently, the preferred direction of movement, is oriented towards the target along the y-axis. In contrast, the participants in Run 030 are not constrained by barriers and form a semi-circle in front of the entrance gate. This leads to a preferred direction of movement with radial orientation towards the target.
The priming procedure had no significant influence on the physical measurands. Therefore, a distinction between "crowding" and "queuing" is omitted when considering these variables. A plot showing the relation between mean density and corridor width in which the priming procedures are labelled is included in [32].

## Density time series

The individual density $\rho_{ind}$ was determined as Voronoi density according to [33] using the software JuPedSim [34, 35]. For each individual, a Voronoi cell is defined which corresponds to the individual's personal space. The individual density is the inverse of the area of the Voronoi cell in terms of persons per square metre.

The density time series shown in Figure 4 are exemplary for all runs and were derived from the mean of the individual densities within the measurement area (cf. Figure 1). With 0.8 m by 0.8 m, the size of the measurement area fits into all corridor widths. To include the area of highest density within the measurement area, its closest border is located 0.5 m from the entrance of the bottleneck.

In the case of high motivation (cf. Figure 4 left), a contraction phase is indicated by a density increase within the first 5 s after the starting signal. For the widest corridor ($w = 5.6$ m), the highest density maximum is observed. For $w = 1.2$ m, the density increase takes longer than for the other runs, indicating the absence of a rapid contraction. It reaches a maximum of 6.5 m$^{-2}$ after ca. 15 s, which is the lowest maximum compared to the other corridor widths.

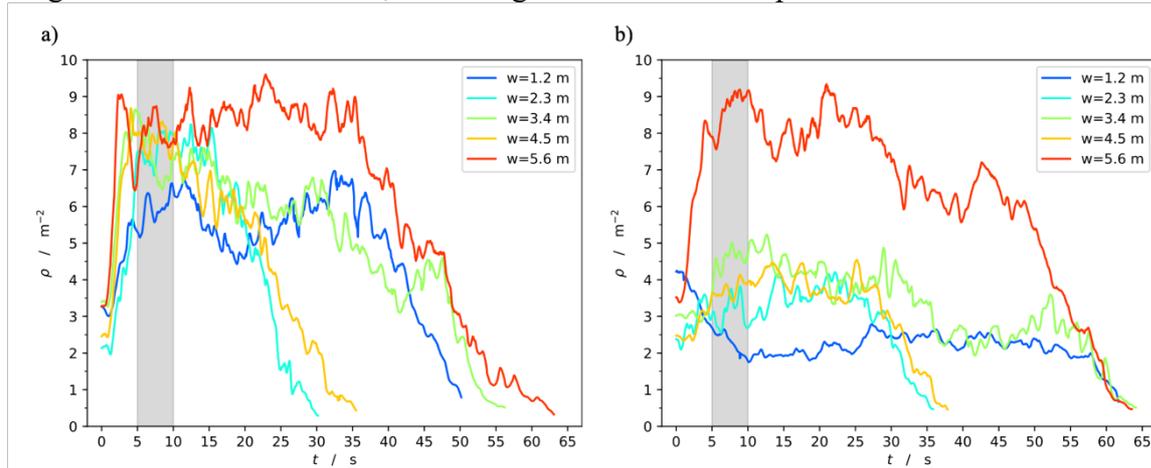

*Figure 4: Time series of the mean density ρ within the measurement area (blue square in Figure 1) for high motivation (a) and low motivation (b). The runs displayed are 110/120 (w = 1.2 m), 230/240 (w = 2.3 m), 270/280 (w = 3.4 m), 050/060 (w = 4.5 m) and 030/040 (w = 5.6 m).*

For lower motivation (cf. Figure 4 right), the maximum of the density time series is generally lower than that for high motivation. This applies to all runs except for the widest corridor where the density maximum is similar to that for high motivation. For the smallest corridor ($b$ = 1.2 m), however, the density decreases from an initial density of 4 m$^{-2}$ to a level of only ca. 2 m$^{-2}$. The slope of the density decrease at the end of the runs seems to be similar for all datasets.

## Relation between density and corridor width

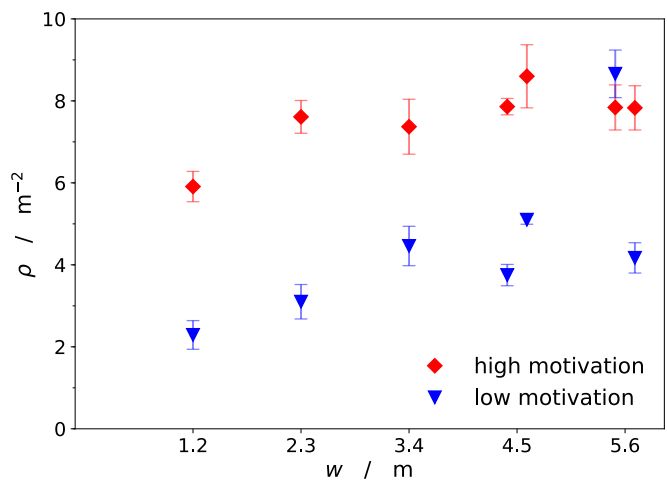

*Figure 5: Mean density ρ within time interval t = 5 s to t = 10 s and within the measurement area versus corridor width w for all runs with N < 40.*

In order to summarise the relation between density and corridor width for all experimental runs, we focus in Figure 5 on the mean value of the density time series within the 5 s interval from $t$ = 5 s to $t$ = 10 s (grey area in Figure 4). This interval is assumed to begin after the initial contraction and to be short enough to end before the density decreases and to be less dependent on variations over time. We chose this interval with the intention of capturing the phase in which the highest density occurs, it is not meant to represent the complete density time series.

We can derive two main findings from Figure 5. First, the mean density increases with increasing corridor width. This applies to both degrees of motivation.
The second and more prominent finding is that the degree of motivation has a large impact on the mean density; that is, for each pair of runs, the mean density is 3-4 m$^{-2}$ higher for the highly motivated run. In other words, there appear to be two discrete density states, depending on the degree of motivation leading to the density gap. It should be noted that the latter finding does not hold for one of the groups in the widest corridor: here, the density for the low motivated run is higher than that of the highly motivated run.

## Waiting time

The waiting time is the remaining time until an individual reaches the target. The target distance is the direct distance to the target along a straight line. In the waiting time plots, both values are plotted against each other for every time step of the experimental run. Each line represents the relation between both values during the runs for one individual. Figure 6 summarises waiting time plots for a representative set of runs. Colour coded at the individual initial positions is the initial velocity within the first 4 s. For individuals being outside of the camera view at $t = 0$ s the initial velocity could not be measured. So, the coloured dot is omitted. The waiting time plots are characteristic of each geometry and motivation.

We can generally divide the waiting time plots into two phases as indicated in Figure 7. The initial phase covers the first 4-5 s after the starting signal (cf. Figure 7-a). In this phase, two opposite behaviours can be observed. Either the participants rush and contract, which is indicated by high initial velocities and lines that are more or less parallel to the x-axis, meaning that they cover a large distance in a short time. Or the participants wait, meaning that they have a small initial velocity and the lines are more or less parallel to the y-axis (see, for example, Run 120). A rushing and contraction is most prominent in the highly motivated run with the widest corridor (Run 030), but it can also be observed in the other runs with a high degree of motivation. The presence of an initial rushing is supported by high initial velocities. Waiting behaviour in the initial phase is most prominent in Run 120, which is a run with a narrow corridor and a low degree of motivation. In some cases, both behaviours occur consecutively within the initial phase, e.g., in Run 280 ($w = 3.4$ m, low motivation).

The congested phase describes the phase in which all interspaces are filled and the participants are in a jam. In order to compare the propagation of individuals towards the target we fit a potential function in the form
$$WT = a \cdot TD^b \quad \text{(i)}$$
to the congested phase of the waiting time plots with $WT$ being the waiting time, $TD$ is the distance to the target, $a$ and $b \in [1, 2]$ are positive parameters. Zuriguel et al. [36] proposed that, the waiting time of an individual is proportional to the number of people who are closer to the target when the flow through the bottleneck and the density are constant. Therefore, the waiting time is proportional to the area $\frac{1}{2}\pi \cdot TD^2$, when the participants are arranged in a semi-circle in front of the bottleneck [36], which is possible in a wide corridor. So, we expect $b \approx 2$ for a corridor with $w = 5.6$ m. For the narrowest corridor with $w = 1.2$ m, we expect that the waiting time is proportional to the target distance ($WT \propto TD^1$) meaning $b \approx 1$. The smaller the corridor, the less space there is to the sides and a semi-circle cannot be formed.

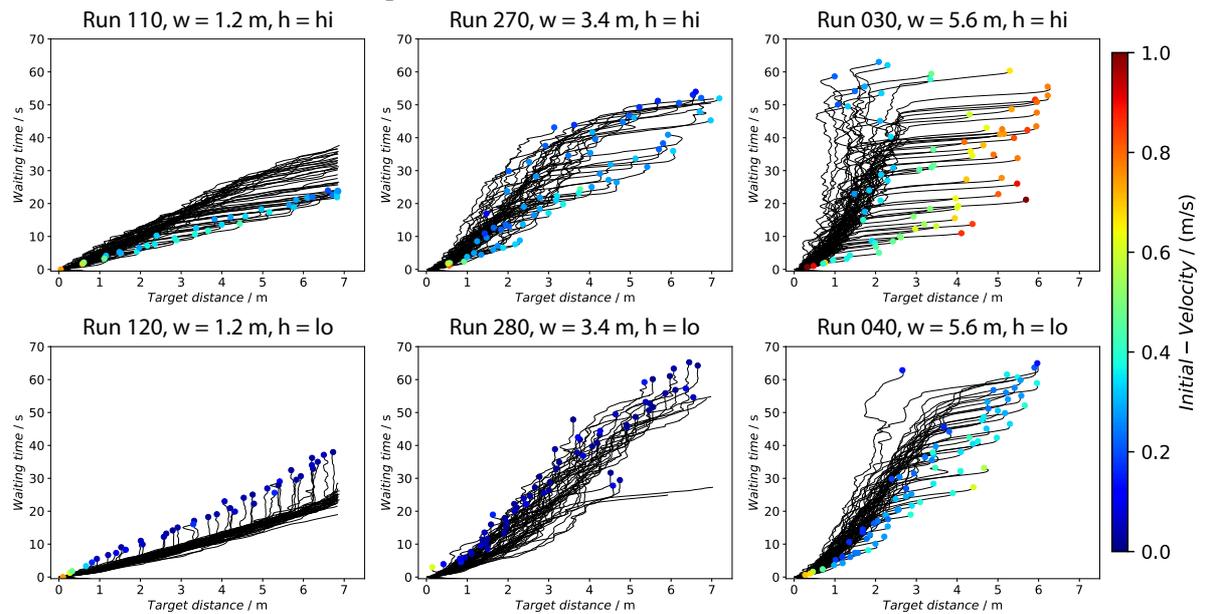

*Figure 6: Individual waiting time versus target distance. The coordinates of the dots indicate the complete runtime and the initial distance to the target (see Figure 1 for target location). The colour of the dots indicates the initial velocity within the first 4 s after the starting signal.*

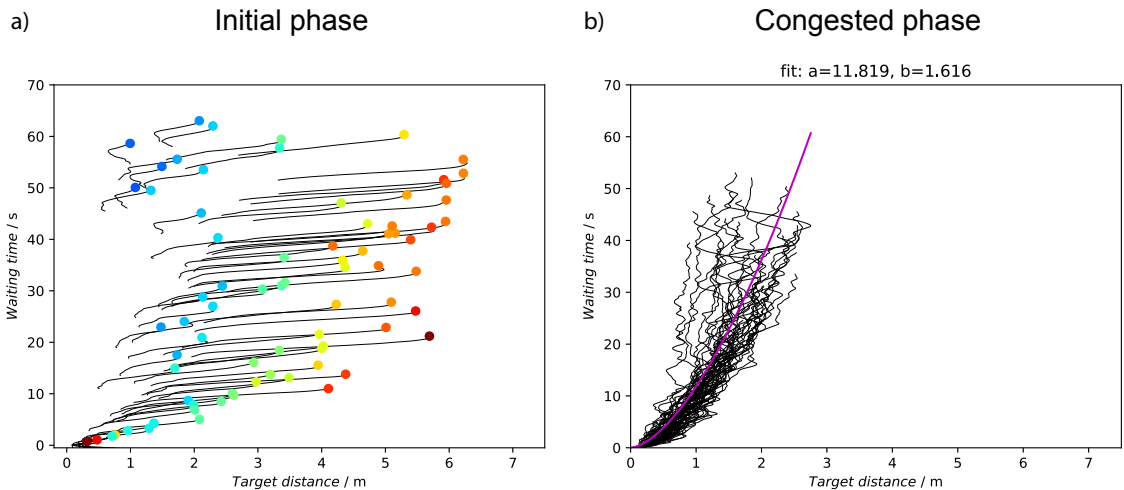

*Figure 7: Two phases of the waiting time plots for Run 030. (a) Initial phase (0-5 s after the starting signal) and (b) congested phase (10 s after the starting signal).*

So, for corridor width between $w = 1.2$ m and $w = 5.6$ m we expect a transition of b from 1 to 2.

An overview over the fit parameters $a$ and $b$ is shown in Table 2. In order to reduce the influence of very slow starters, the first 10 s of each run were omitted. As expected, $b$ is closest to 1 in the narrowest corridor. It increases with increasing corridor width. However, the maximum value $b = 1.6$ is lower than the expected $b \approx 2$. It is assumed that the trend will be more pronounced for experiments with larger numbers of participants.

To estimate a measure of the velocity towards target direction, we build the first partial derivation of equation (i) with respect to *TD* yielding

$$\frac{\partial WT}{\partial TD} = (a \cdot b)TD^{(b-1)} \quad \text{(ii)}$$

The inverse $(a \cdot b)^{-1}$ has the unit of a velocity and is summarized in Figure 8. The wider the corridor and the higher the degree of motivation, the smaller $(a \cdot b)^{-1}$ and the slower the propagation towards the target. This closely correlates with the high densities found for wide corridor width and high degree of motivation.

Table 2: Parameters a and b resulting from a fit of a potential function $WT = a \cdot TD^b$ to the waiting time plots omitting the first 10 s after the starting signal.

|  |  | high motivation | | low motivation | |
| --- | --- | --- | --- | --- | --- |
| **Run** | **Width $w$ / m** | a | b | a | b |
| 110 / 120 | 1.2 | 5.19 | 1.04 | 2.21 | 1.24 |
| 230 / 240 | 2.3 | 7.06 | 1.19 | 4.26 | 1.31 |
| 270 / 280 | 3.4 | 7.50 | 1.26 | 4.39 | 1.47 |
| 050 / 060 | 4.5 | 8.99 | 1.62 | 5.13 | 1.41 |
| 250 / 260 | 4.5 | 7.58 | 1.27 | 4.86 | 1.36 |
| 030 / 040 | 5.6 | 11.82 | 1.62 | 8.50 | 1.45 |
| 150 / 160 | 5.6 | 11.06 | 1.30 | 5.65 | 1.27 |

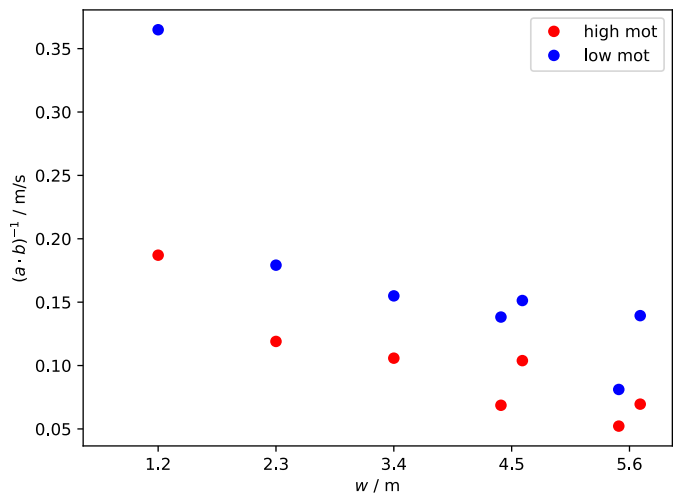

Figure 8: Fit parameters $(a \cdot b)^{-1}$ against corridor width.

## Flow through the bottleneck

To evaluate the flow through the bottleneck, we look at the time gaps $\Delta t$ between consecutive persons crossing the target line located within the entrance gate. Shorter time gaps represent a higher flow through the bottleneck. In order to exclude boundary effects, we omit the first and last ten persons crossing the target line.

Figure 9 shows boxplots of $\Delta t$ for each run with more than 40 participants. Apparently, the corridor width has no influence on $\Delta t$. But for most experimental pairs, $\Delta t$ is longer for a low degree of motivation and shorter for the corresponding run with a high degree of motivation. This means that we did not find any evidence of the 'faster-is-slower' effect [13]. In contrast, our findings suggest that the flow through the bottleneck is not necessarily low for higher motivation but rather the other way round. However, this only applies to small to intermediate corridor width between $w = 1.2$ m through $w = 3.4$ m. For wider corridors of $w = 5.4$ m and $w = 5.6$ m, the dispersion of the data around the mean value is too large to derive a clear statement.

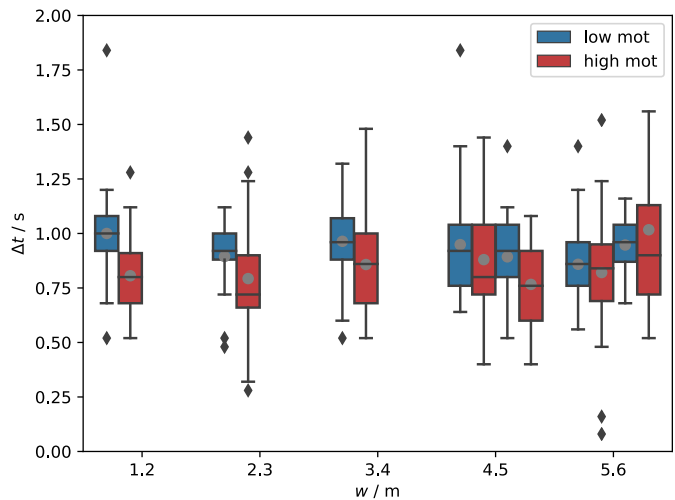

*Figure 9: Boxplot representation of time gaps Δt between consecutive persons crossing the target line within the bottleneck for each run with N>40. The first and last 10 people crossing the target line were omitted. Grey dots: mean values; diamonds: outliers.*

# Results: questionnaire

## Experimental conditions: width and priming

On ratings of fairness, both width ($F(9;468)=5.376$, $p=.000$) and priming ($F(9;468)=9.904$, $p=.001$) have a significant effect. Their interaction effect is also significant ($F(9;468)=4.091$, $p=.004$). Because Levene's test shows that variances are not equal, the Welch-James test is used for this analysis. Post hoc tests (for unequal variances) reveal that participants in the priming condition "crowding" rate the procedure as significantly fairer (M=3.08, SD=1.30, $p=.000$) than those in the priming condition "queuing" (M=2.55, SD=1.45). Furthermore, the width of 1.2 m (M=3.33, SD=1.34) is rated as being significantly fairer than 2.3 m (M=2.61, SD=1.44, $p=.016$), 4.5 m (M=2.61, SD=1.26, $p=.002$) and 5.6 m (M=2.71, SD=1.42, $p=.006$), but not as 3.4 m (M=2.98, SD=1.31, $p=.468$). Both effects are illustrated in Figure 10.

On the rating of comfort, only width has a significant effect ($F(9,470)=4.15$, $p=.002$). Again, the Welch-James test is performed for this analysis because of unequal variances. Post hoc tests (for unequal variances) show that the entrance procedure with 1.2 m (M=3.4, SD=1.26) is rated as being significantly more comfortable than 4.5 m (M=2.82, SD=1.31, $p=.021$) and 5.6 m (M=2.77, SD=1.34, $p=.002$). Although the rating in the priming condition "crowding" looks more positive, this trend is not significant ($F(9,470)=2.96$, $p=.154$). There is no significant interaction effect, either ($F(9,470)=1.25$, $p=.297$). These effects and trends are illustrated in Figure 11.

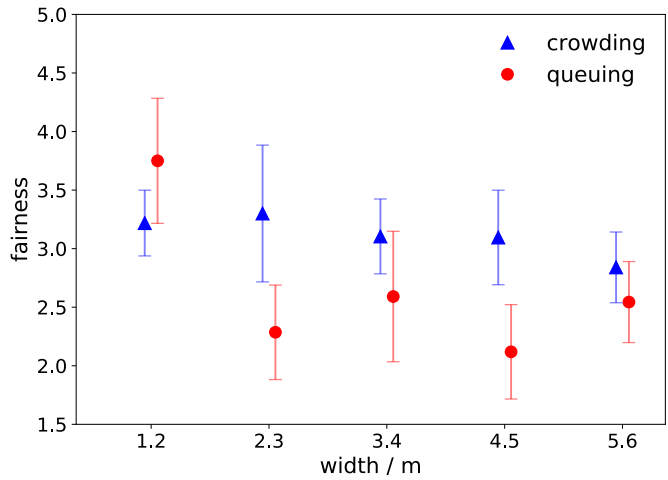

*Figure 10: Ratings of fairness for width and priming (error bars: 95% CI).*

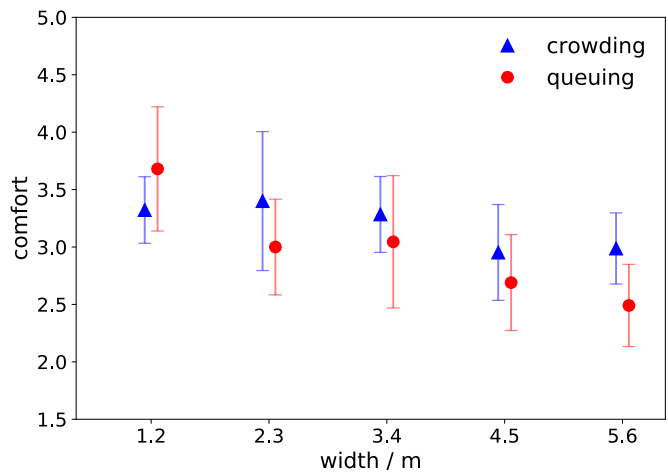

*Figure 11: Ratings of comfort for width and priming (error bars: 95% CI).*

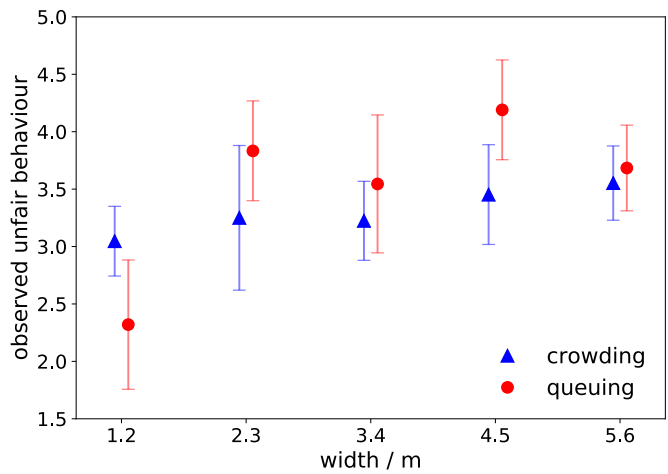

*Figure 12: Observation of unfair behaviour for width and priming (error bars: 95% CI).*

Width has a significant effect on the observation of unfair behaviour ($F(9, 469)=7.48$, $p=.000$). In this case, tests for equal variances can be used. Bonferroni post hoc tests show that in the entrance procedure with 1.2 m, participants observe significantly less unfair behaviour (M=2.88, SD=1.41) than in the ones with 2.3 m (M=3.65, SD=1.55, p=.009), 4.5 (M=3.82, SD=1.50, p=.000) or 5.6 m (M=3.61, SD=1.42, p=.001). Priming does not have a significant effect (F(9, 469)=2.02, p=.156). Only a non-significant trend for priming can be observed: in "crowding", less unfair behaviour is observed. This is in line with the results for fairness and comfort. The interaction effect of priming and width is also significant (F(9, 469)=3.05, p=.017) and illustrated in Figure 12. Comparing Figure 10 and 12 it becomes visible that these ratings are negatively correlated (a procedure is less fair if more people show unfair behaviour; r=-.452, p=.000). Ratings of comfort are also correlated with ratings of fairness (r=.466, p=.000) and unfair behaviour (r=-.423, p=.000).

Width does not affect the percentage of participants who think that they can contribute to accessing faster ($\chi^2(4)$=1.13, p=.890). Nor does it significantly affect the percentage of participants who note that they themselves pushed ($\chi^2(4)$=6.77, p=.149). Table 3 shows the percentage for contribution and pushing for all five widths. Participants in 1.2 m appear to claim that they were engaged in pushing less frequently, but this effect is not significant.

*Table 3: Percentage of participants who think that they can contribute or who say that they did push.*

|  | 1.2 m | 2.3 m | 3.4 m | 4.5 m | 5.6 m |
|---|---|---|---|---|---|
| Contribution to accessing faster (yes) | 58.9% | 62.9% | 64.0% | 59.5% | 57.9% |
| Engaged in pushing (yes) | 46.8% | 64.5% | 58.4% | 58.3% | 60.2% |

In the priming condition "queuing", significantly more participants say that they did engage in pushing ($\chi^2(1)$=4.67, p=.031): 63.1 % in comparison to 53.1% in the priming condition "crowding". Furthermore, 64.9% of participants in the queuing condition and 57.2% in the crowding condition think that they were able to contribute to accessing faster. But this difference does not reach the level of significance ($\chi^2(1)$=2.83, p=.092).

To sum up, the expected effects of width can only be confirmed for 1.2 m: 1.2 m is rated as fairer and more comfortable. Less unfair behaviour is observed in this condition. Furthermore, a smaller percentage of participants tend to say that they did engage in pushing themselves, but this effect is not significant. Priming has significant effects on fairness and pushing behaviour – and these effects are contrary to our hypotheses.

## Post hoc analysis: percentage of females in the group

There are differences between the experimental runs which are unaccounted for by the experimental conditions alone. For that reason, we add an explorative analysis of the percentage of females in the run. A combination of this factor with width or priming in a single GLM is not possible because "percentage of females" as a post hoc created variable is not systematically varied along with the other two variables (resulting in missing data for several combinations of variables). Overall, 46.9% females and 52.2% males participated.

However, runs vary in percentage of females from 13.6% to 87.3%. A categorical variable was created for 1:1-33.3% females, 2:33.4-66.6% females and 3: 66.7-100% females. The percentage of females significantly affects ratings of fairness ($F(2,475)=6.91$, $p=.001$) and unfair behaviour ($F(2,476)=5.94$, $p=.003$) but not of comfort ($F(2,477)=2.19$, $p=.113$). Groups with 1-33.3% female are rated as being less just (M=2.65, SD=1.45) than groups with 33.4-66.6% females (M=3.21, SD=1.36, p=.001). Also, in groups with 1-33.3% female, more unfair behaviour (M=3.63, SD=1.49) is observed than in groups with 33.4-66.6% females (M=3.05, SD=1.56, p=.002). No other differences reach the level of significance. To sum up, runs with markedly more men are rated as being less just and as showing more unfair behaviour than runs with approximately equal percentages of men and women.

### Individual differences

Furthermore, the following two gender differences in ratings can be observed: males rate the entrance procedures as more comfortable (M=3.20, SD=1.47) than females (M=2.89, SD=1.30 p=.019). Significantly more men than women (62.4% versus 51.6%) say that they did engage in pushing ($\chi^2(1)=5.69$, $p=.017$). On the other scales, no significant gender differences occur.

Those who state that they are able to contribute to accessing faster rate the procedure as being more comfortable (M=3.18, SD=1.41) than the others (M=2.87, SD=1.37, p=.019). Those who say that they did engage in pushing themselves rate the entrance procedure as being less just (M=2.69, SD=1.29) than the others (M=3.12, SD=1.47, p=.001) and observe more unfair behaviour (M=3.64, SD=1.39) than the others (M=3.15, SD=1.54, p=.000).

### Qualitative results

Answers on open-ended questions were slightly generalised and categorised, translated and counted. Here, only the four most frequently used categories are reported. Since participants were allowed to give several answers, the total number of answers does not add up to the total number of participants.

*What rules applied to the entrance procedure?*
Unfortunately, this question did not work well in the questionnaire. Only 31 participants filled in the blank space. Most often the rule "survival of the fittest" is named (19). Interestingly, "survival of the fittest" is more often listed in the priming condition "queuing", particularly for 2.3, 4.5 and 5.6 m (Run 230, 250, 150). These runs show high mean densities (see above). Furthermore, five participants say that one should take care of others. Two participants list "zip merging" and two "do not push hard". The norm of queuing is not mentioned at all.

*What kind of unfair behaviour did you observe?*
Pushing forward is mentioned most often as a form of unfair behaviour (535 answers), followed by use of elbows (23), lack of consideration for others (20) and assertion of the strongest (13).

*What strategies did you use for accessing faster/Why could you not use strategies for accessing faster?*

Most often, pushing forward is named as a strategy (131), followed by filling gaps (57), choosing the right or left side or middle as the preferred pathway (46) and standing at the front at the beginning (37).
The following reasons for not being able to influence the entrance procedure were given: no room for manoeuvre (100), an initial position at the back (30), being too weak (16), consideration for others (9).

*Why did you yourself push /Why did you yourself not push?*
Most participants say that they pushed because they were instructed to be fast or one of the first (182). Others say that they were pushed from the back (16), were trying to escape (14) or pushed because they saw others pushing (14).
Participants say they did not push because they have an aversion to pushing (34), out of consideration for others (28), because they wanted to avoid danger (26) or because pushing is inefficient.

## Combined results

In order to combine physical and psychological results, we correlate mean density and initial velocity with ratings of fairness, comfort and unfair behaviour (cf. Tables 4-5). Since runs with less than 40 participants are too short to determine comparable physical characteristics, only those seven runs with more than 40 participants are included into the analysis. The following correlations can be observed: the higher mean density is, the more unfair behaviour is observed (r=.787, p=.036) and the faster initial velocity is, the more unfair behaviour is observed (r=.768, p=.044). Furthermore, the following meaningful, though not significant, correlations occur. Runs with higher mean density are rated less just and less comfortable. Runs with faster initial velocity are also rated as less just and less comfortable.

*Table 4: Correlations (Pearson's r) with mean density.*

|  | Mean fairness | Mean comfort | Mean unfair behaviour |
|---|---|---|---|
| Mean density (N=7) | -.643 | -.651 | .787* |
|  | P=.119 | P=.113 | P=.036 |

*Table 5: Correlations (Pearson's r) with mean initial velocity (0-4 s).*
*\*Correlation is significant at the 0.05 level (two-tailed).*

|  | Mean fairness | Mean comfort | Mean unfair behaviour |
|---|---|---|---|
| Initial velocity (N=7) | -.714 | -.695 | .768* |
|  | P=.072 | P=.083 | P=.044 |

The effects of priming and gender of participants on density are calculated with multiple correlation coefficients (together with width and motivation). No significant correlations for priming or gender can be observed.

# Discussion

## Variable number of participants

The recruitment method resulted in varying numbers of participants (see Table 1). Runs with less than 40 participants had such a short duration that physical mean quantities are not meaningful. Furthermore, the number of participants became an influential factor itself which we were unable to systematically vary and control. Future research should choose a more controllable recruitment method and include the number of participants as an experimental factor in itself.

## Effects of width and motivation on density, waiting times, velocity and flow

According to our findings, the density in front of the bottleneck increases with increasing corridor width. We assume that this results from boundary effects and a larger space offering more potential directions to move and fill gaps. Furthermore, we found two density levels dependent on the degree of motivation for each width, meaning that the density is about four people per square metre higher when the participants are highly motivated.

Regarding the velocity, it was found that a high degree of motivation leads to high initial velocity but low velocity in the target direction during the congested phase, which is due to a high density after the initial contraction phase. In the corresponding less motivated runs, the initial velocity is lower and the velocity during the congested phase is slightly higher than in the corresponding highly motivated runs, which is in turn a consequence of the lower density. Overall, the velocity in target direction is slower, the wider the corridor is. This is a consequence of the geometry in which the participants can arrange themselves in front of the bottleneck depending on the space provided by the barriers: in a narrow corridor, the participants are forced into a 'queue-like' arrangement. Everyone is oriented along the y-axis and so is their preferred direction of movement. Directly at the entrance of the bottleneck, only two people at a time compete to enter (zipper method). In contrast, in a wide corridor, the participants have more space to form a semi-circle. They are closer to the target earlier but their orientation is in radial direction towards the target. The result is that more people compete to enter the bottleneck, which results in more conflicts than in the zipper method. For intermediate corridor width, we find a transition between both situations.

For the fit of a potential function to the waiting time plots during the congested phase it was expected that the exponent $b$ is close to one for a very narrow corridor because the number of people in front is proportional to the distance to the target and that the parameter $b$ is close to two for a wide corridor because the number of people in front is proportional to the area of a semi-circle. These expectations were only partially fulfilled. For the smallest corridor, the exponent $b$ is close to one and $b$ increases the wider the corridor becomes, but it does not reach the expected value of two. Several reasons could be the cause: even if the corridor is wide enough, the participants do not necessarily form a semi-circle in front of the bottleneck. Also V-shaped arrangements or queues occur, especially, when the motivation is low. Furthermore, the density is not constant and the number of participants is finite and varies.

According to the time gaps of consecutive persons crossing the target line, a high degree of motivation results in a higher flow through the bottleneck. This means that highly motivated people enter more quickly and we can therefore observe a faster-is-faster effect. It is only for the widest corridor that the effect appears to be reversed. We note that these findings only hold for the bottleneck width of 0.5 m that only one person can cross at a time. In other experiments with high motivation at bottlenecks, a reduced flow in comparison to a low degree of motivation could be observed [16]. But since the width of the bottleneck that was used in the experiments by [16] was wider than in our experiment (allowing more than one person at a time to enter the bottleneck), the effects are not fully comparable. As outlined in the introduction, the faster-is-slower effect has only a weak experimental evidence and is not reproduced by our experiments. A more systematic analysis of the relationship between motivation and flow through narrow bottlenecks is necessary.

Furthermore, it is notable that we found no significant dependence of the flow on corridor width.

Statistical tests concerning the effects triggered by corridor width and motivation on density, velocity and flow were not included in this manuscript because of the limited number of samples. Note that we also did not vary the ordering of the motivation treatment (always high motivation first). Future research could test whether a reverse order changes the outcomes. We assume, however, that the instructions induce high and low motivation independent from their ordering. The strong effects of the instructions are visible in the videos.

## Queuing not relevant

None of the corridor widths are associated with the social norm of queuing. Our initial hypothesis that narrower corridors trigger the norm of queuing was not confirmed. Neither did priming make queuing more salient. It can therefore be concluded that the norm of queuing is not a relevant factor in the emergence of lower densities in narrower corridors. The discrepancy from our earlier study (in which queuing appeared to be a relevant factor) might be related to two methodological issues: in the earlier study [18] the questionnaire was filled in after seeing the videos of the entrance procedure from a bird's eye perspective. Perhaps the dynamics in narrow corridors looks like a queue from above but not from the perspective of a participant in the crowd. Also, the question "what rules applied to the entrance procedure?" did not work well in the study described here: only 31 participants (out of 479) answered. Maybe the question was not properly understood and we therefore were unable to detect the norm of queuing.

Priming did not affect density and induced a contrast effect on some of the ratings: when primed with the norm of queuing, runs were evaluated as less just and as showing more unfair behaviour. This might be explained as follows: priming does not change behaviour but influences participants' expectations of the entrance procedure. The priming "queuing" makes participants expect an orderly procedure. The fact that the real entrance procedure diverges from this expectation is then reflected in more negative ratings. Interestingly, for w=1.2 m, priming showed the expected effects: after the priming queuing the procedure was rated as being fairer and as showing less inappropriate behaviour. Maybe the procedure with 1.2 m looked more like a queue and therefore met the expectations of the participants. Future research might have to use a priming procedure which is less conceptual but more behavioural.

Furthermore, we did not use any methodical tool in order to register different group formations. We did observe that experimental runs were different – showing more or less social cohesion and different degrees of social identification. Future experiments should make these factors measurable in order to account for their effects on density, velocity or ratings of the entrance situation – as shown in research on psychological and physical crowds [37, 38].

### Ratings of entrance situations

The narrow corridor with 1.2 m is rated as more just, more comfortable and as showing less unfair behaviour. These results can potentially be interpreted in the following way: since it is possible to oversee the corridor, the FIFO principle can be controlled and therefore, the procedure appears to be more just. Because in corridors with 1.2 m lower densities also occur, they might be rated as more comfortable.

The correlation between density and observed unfair behaviour can be interpreted in two ways: either high densities result from pushing behaviour which is evaluated as unfair or participants observe more unfair behaviour and then start pushing and thus increase density. The correlation between initial velocity and observed unfair behaviour can be interpreted similarly: either participants evaluate other participants who are moving fast as behaving unfairly or because they expect unfair behaviour by others, they start moving fast themselves at the beginning.

This study confirms that pushing is normatively ambivalent: it is at the same time listed as inappropriate behaviour and as a strategy (by the same participants). Furthermore, individual attitudes about pushing are diverse: approximately half of the participants say that they were engaged in pushing themselves and the other half not. Some participants articulate their concerns about pushing being dangerous, rude or inefficient. Others do not see it as a problem. This ambiguity of pushing behaviour might be quite influential in crowd dynamics: if, culturally, pushing is neither clearly forbidden nor seen as clearly legitimate, it might occur in a context-dependent and dynamic way. In this sample, the percentage of participants saying that they did push varies from 29.2% to 78.6%. However, because we only have questionnaire data for one point in time, we cannot track any dynamical changes in pushing being seen as more or less appropriate in the course of a run.

It is interesting to note that pushing is not the only strategy listed. Furthermore, filling gaps, using a specific path (left, right or middle) and standing at the front before the entrance opens are other strategies mentioned. Using gaps is of particular interest because it increases density as does pushing but it is probably seen as being more considerate. Future research should focus on pushing as well as the behaviour of filling gaps.

## Acknowledgements

We are grateful for the remarks of the unknown referees and the editor. Especially the idea to fit a potential function to the waiting time plots was a valuable contribution to the analysis presented in this manuscript.
We would like to thank all students of the University of Wuppertal, Germany for taking part in the experimental study. Furthermore, we thank all our colleagues who took part in planning and implementing the experiments.

# Ethics

Full ethical clearance for this research was granted by the ethics board at the University of Wuppertal, Germany. Written and informed consent was obtained from all participants who took part in the experiment and in the consecutive questionnaire study.

# Data

All raw data, i.e. video recordings, head trajectories and questionnaire answers, are available through the Pedestrian Dynamics Data Archive hosted by the Forschungszentrum Juelich, https://doi.org/10.34735/ped.2018.1.

# Funding

# Competing interests

The authors state no competing interests that could affect the objectivity of this research.

# Authors' contributions

All authors contributed equally to this work.
All authors gave final approval for publication and agree to be held accountable for the work performed therein.